\documentclass[aps,pre,twocolumn,groupedaddress,showpacs,amssymb,amsmath]{revtex4-1}
\usepackage{prettyref}
\usepackage{graphicx}
\usepackage{verbatim}   
\usepackage{color}      
\usepackage{subfigure}  
\usepackage{hyperref}
 {%
\begin{document}


\title{A generalized Caroli formula for transmission coefficient with lead-lead coupling}

\author{Huanan Li}

\email{g0900726@nus.edu.sg}


\affiliation{Department of Physics and Center for Computational Science and Engineering,
National University of Singapore, Singapore 117542, Republic of Singapore }

\author{Bijay Kumar Agarwalla}

\affiliation{Department of Physics and Center for Computational Science and Engineering,
National University of Singapore, Singapore 117542, Republic of Singapore }

\author{Jian-Sheng Wang}

\affiliation{Department of Physics and Center for Computational Science and Engineering,
National University of Singapore, Singapore 117542, Republic of Singapore }

\date{28 May 2012}
\begin{abstract}
We present a generalized transmission coefficient formula for the
lead-junction-lead system, in which interaction between the leads has
been taken into account. Based on it the Caroli formula
could be easily recovered and a transmission coefficient formula for
interface problem in the ballistic system can be obtained.
The condition of validity for the formula is carefully explored.
We mainly focus on heat transport.  However, the
corresponding electrical transport could be similarly dealt with.
Also, an illustrative example is given to clarify the precise meaning
of the quantities used in the formula, such as the concept of the
reduced interacting matrix in different situations. In addition, an explicit
transmission coefficient formula for a general one-dimensional interface setup is
obtained based on the derived interface formula.
\end{abstract}

\maketitle

\section{INTRODUCTION\label{sec:INTRODUCTION}}

In recent years there has been a huge increase in research and
development of nanoscale science and technology, with the study of
energy and electron transport playing
important role. Focusing on thermal transport, Landauer-like results
for steady-state heat flow have been proposed earlier \cite{Rego1998,Blencowe1999}.
Subsequently, based on quantum Langevin equation approach, many authors
successfully obtained a Landauer-type expression \cite{Segal2003,Dhar2006,Dhar2008}.
Alternatively, nonequilibrium Green's function (NEGF) method has been
introduced to investigate mesoscopic thermal transport, which is particularly
suited for use with ballistic thermal transport and readily allows
the incorporation of nonlinear interactions \cite{Ozpineci2001,Wang2006,Galperin2007}.
Generally speaking, in the lead-junction-lead system, steady-state
heat current of ballistic thermal transport flowing from left lead
to right lead has been described by the Landauer-like formula, which
was derived first for electrical current, as
\begin{alignat}{1}
I= & \int_{0}^{\infty}\frac{d\omega}{2\pi}\,\hbar\omega \,T\left[\omega\right]\left(f_{L}-f_{R}\right),
\end{alignat}
where $f_{\left\{ L,R\right\} }=\left\{ \exp\left({\hbar\omega}/{k_{B}T_{\left\{ L,R\right\} }}\right)-1\right\} ^{-1}$
is the Bose-Einstein distribution for phonons, and $T\left[\omega\right]$
is known as the transmission coefficient. Based on nonequilibrium
Green's function method, $T\left[\omega\right]$ can be calculated
through the Caroli formula in terms of the Green's functions of the junction
and the self-energies of the leads,
\begin{alignat}{1}
T\left[\omega\right]= & \mathrm{Tr}\left(G^{r}\Gamma_{R}G^{a}\Gamma_{L}\right),\label{eq:Caroli formula}
\end{alignat}
where $G^{r,a}$ is the Green's function of the junction, and
\begin{alignat}{1}
\Gamma_{\left\{ L,R\right\} }= & i\left[\Sigma_{\left\{ L,R\right\} }^{r}-\Sigma_{\left\{ L,R\right\} }^{a}\right],
\end{alignat}
where the self-energy terms $\Sigma^{r,a}_{\left\{ L,R\right\} }$ are
due to the semi-infinite leads on the left, $L$, and on the right,
$R$, respectively. The superscript $r$ and $a$ denote the retarded
and advanced, respectively, both for the self-energies as well as for
the Green's functions in the formula. The specific form \prettyref{eq:Caroli formula}
was given from NEGF formalism by Meir and Wingreen \cite{Meir1992}
for electronic case and later by Yamamoto and Watanabe for phonon transport \cite{Yamamoto2006},
while Caroli et al. first obtained a formula for the electronic transport
in a slightly more restricted case \cite{Caroli1971}. Also, Mingo
et al. have derived a similar expression for transmission coefficient
using ``atomistic Green's function'' method \cite{Mingo2003,Zhang2007}. Very recently,
Das and Dhar \cite{Suman2012} derive the Landauer-like expression
from plane wave picture using Lippmann-Schwinger scattering approach.

The Landauer-like formula describes the situation in which the junction
is small enough compared to the coherent length of the waves so
that it could be treated as elastic scattering where the energy
is conserved. Furthermore, it has been assumed that the two leads
are decoupled which physically means there is no direct tunneling
between the two leads. Through modern nanoscale technology, small
junction is easily realized such as in certain nanoscale systems,
for instance, a single molecule or, in general, a small cluster of
atoms between two bulk electrodes. In that case, the electrode surfaces
of the bulk conductors may be separated by just a few angstroms so
that some finite electronic coupling between the two surfaces is inevitable
taking into account the long-range interaction. In order to solve
this problem, Di Ventra suggested that \cite{Ventra2008} we can choose our ``sample''
region (junction) to extend several atomic layers inside the bulk
electrodes where screening is essentially complete so that the above
coupling could be negligible. It turns out to be
correct using this trick to avoid the interaction between the two
leads, which will be verified in a simple example at the end of the
paper, even though we, to some limited extent, modify the initial
condition necessary to derive Landauer-like formula in NEGF formalism
and repartition the total Hamiltonian. However, this procedure or
trick could not be always done due to some topological reason such
as studying heat current in Rubin model \cite{Rubin1971} in which the other end of
the two semi-infinite leads is connected (a ring problem). Actually
this somewhat trivial example is not so artificial since it is equivalent
to using periodic boundary condition in Rubin model. Furthermore,
the modification of the initial product state will certainly affect
the behavior of the transient heat current. If we want to study the
transient and steady heat current \cite{Bijay2011} in a unified way, the repartitioning
procedure which changes the model is not acceptable. So in this work
we will try to derive a compact formula applicable to this general
model including lead-lead interaction for steady-state heat current
according to NEGF formalism, and correspondingly obtained a Caroli-like
formula for transmission coefficient. Furthermore, an interface transmission
coefficient formula in the NEGF formalism will be given as a special
case of the general Caroli-like formula. Also, the standard Caroli
formula follows as a one-line proof.

The paper is organized into two main sections. In \prettyref{sec:FORMALISM},
we develop our formalism to derive our generalized expression for
the steady current directly taking coupling between leads into account.
Based on this general formula, we will recover the Caroli formula
and derive a computationally efficient interface formula in \prettyref{sub:Recovering-Caroli-formula}.
Then we apply this formalism to an illustrative model system in
\prettyref{sec:AN-ILLUSTRATIVE-APPLICATION} and show the results
of numerical calculations. Also, we will apply the interface formula obtained in \prettyref{sec:FORMALISM}
to derive an explicit expression for transmission coefficient in \prettyref{sec:AN-Interface-APPLICATION}. Finally we conclude with a short discussion in Section~V.

\section{FORMALISM\label{sec:FORMALISM}}

\subsection{Model system\label{sub:Model-system}}

As was mentioned previously, we will consider the lead-junction-lead
model initially prepared in product state $\hat{\rho}\left(t_{0}\right)=\frac{e^{-\beta_{L}H_{L}}}{\mathrm{Tr}\left(e^{-\beta_{L}H_{L}}\right)}\otimes\frac{e^{-\beta_{C}H_{C}}}{\mathrm{Tr}\left(e^{-\beta_{C}H_{C}}\right)}\otimes\frac{e^{-\beta_{R}H_{R}}}{\mathrm{Tr}\left(e^{-\beta_{R}H_{R}}\right)}$.
We can imagine that left lead $\left(L\right)$, center junction $\left(C\right)$,
and right lead $\left(R\right)$ in this model was in contact with
three different heat baths at the inverse temperature $\beta_{L}\equiv\left(k_{B}T_{L}\right)^{-1}$,
$\beta_{C}\equiv\left(k_{B}T_{C}\right)^{-1}$ and $\beta_{R}\equiv\left(k_{B}T_{R}\right)^{-1}$,
respectively for time $t<t_{0}$. At time $t=t_{0}$, all the heat
baths are removed, and coupling of the center junction with the leads
and the interaction between the two leads are switched on abruptly.
Now the total Hamiltonian of the lead-junction-lead system becomes
\begin{align}
H_{tot}= & H_{L}+H_{C}+H_{R}+H_{LC}+H_{CR}+H_{LR},\label{eq:total H}
\end{align}
where $H_{\alpha}=\frac{1}{2}p_{\alpha}^{T}p_{\alpha}+$$\frac{1}{2}u_{\alpha}^{T}K^{\alpha}u_{\alpha},\;\alpha=L,C,R$
represents coupled harmonic oscillators, $u_{\alpha}\equiv\sqrt{m}x_{\alpha}$
and $p_{\alpha}$ are column vectors of transformed coordinates and
corresponding conjugate momenta in region $\alpha$. The superscript
$T$ stands for matrix transpose. $H_{LC}\equiv u_{L}^{T}V^{LC}u_{C}$
and $H_{CR}\equiv u_{C}^{T}V^{CR}u_{R}$ are the usual couplings between
the junction and the two leads, which are certainly necessary to
establish the heat current. Now the new term representing interaction
between two leads $H_{LR}=u_{L}^{T}V^{LR}u_{R}$ will modify transmission
coefficient greatly, which is our main interest.

It is worth mentioning that nonlinear interaction could be added inside
the center junction and dealt with using self-consistent approach
in the framework of NEGF, which has been done by many authors \cite{Ness2010,Xu2008}.

\subsection{Steady state contour-ordered Green's functions}

Contour-ordered Green's functions are the central objects in the NEGF
formalism, among which the directly derived relation say, Dyson equation,
could be readily transformed to all kinds of relations among the real-time
Green's functions by Langreth theorem \cite{Haug2008}. And many interesting quantities
such as the current we will consider in the following subsection could
be easily related to the proper real-time Green's functions.

Steady-state contour-ordered Green's functions are defined as
\begin{alignat}{1}
G_{jk}^{\alpha\beta}\left(\tau_{1},\tau_{2}\right)= & -\frac{i}{\hbar}\mathrm{Tr}\left\{ \hat{\rho}^{ss}\left(s\right)T_{c}\left[u_{j}^{\alpha}\left(\tau_{1}\right)u_{k}^{\beta}\left(\tau_{2}\right)\right]\right\} ,
\end{alignat}
where $\hat{\rho}^{ss}\left(s\right)=U\left(s,t_{0}\right)\hat{\rho}\left(t_{0}\right)U\left(t_{0},s\right)$
is the steady-state density operator, in which time $s>t_{0}$ introduced
for convenience of later discussion could take any finite time since
the switch-on time $t_{0}$ will be let to go to $-\infty$ at the
end in order to establish steady-state heat current. $u_{j}^{\alpha}\left(\tau_{1}\right)=U\left(s,\tau_{1}\right)u_{j}^{\alpha}U\left(\tau_{1},s\right)$
is operator in the Heisenberg picture and similarly for $u_{k}^{\beta}\left(\tau_{2}\right)$.
The variables $\tau_{1}$ and $\tau_{2}$ are on the contour from
time $s$ to $\infty$ and back from $\infty$ to time $s$. $U\left(t_{0},s\right)$
etc. are the time evolution operators of the full Hamiltonian. $T_{c}$
is the contour-ordering super-operator. There is a strong assumption
here which is all we need in the whole derivation, where we assume
steady state could be established from initial product state after
infinite time so that all the steady-state real-time Green's function
depend only on the difference of the two time arguments. This intuitively
reasonable assumption is not always guaranteed and there is a specific
example about how to establish steady-state heat current in Ref \cite{Cuansing2011}.

After $t_{0}\rightarrow-\infty$ , $s\rightarrow t_{0}^{+}$, and
transforming to the interaction picture, where the total Hamiltonian
$H_{tot}$ is separated into the free part $H_{0}=H_{L}+H_{C}+H_{R}$
and the interaction part $H_{int}=H_{LC}+H_{CR}+H_{LR}$, we obtain
\begin{align}
 & G_{jk}^{\alpha\beta}\left(\tau_{1},\tau_{2}\right) =\nonumber \\
& -\frac{i}{\hbar}\mathrm{Tr}\left\{ \hat{\rho}\left(-\infty\right)T_{c}\left[e^{-\frac{i}{\hbar}\int_{K}H_{int}^{I}\left(\tau'\right)d\tau^{'}}u_{I,j}^{\alpha}\left(\tau_{1}\right)u_{I,k}^{\beta}\left(\tau_{2}\right)\right]\right\} ,
\end{align}
where $u_{I,j}^{\alpha}\left(\tau_{1}\right)=e^{\frac{i}{\hbar}H_{0}\tau_{1}}u_{j}^{\alpha}e^{-\frac{i}{\hbar}H_{0}\tau_1}$
is operator in the interaction picture and similarly for $u_{I,k}^{\beta}\left(\tau_{2}\right)$ and
$H_{int}^{I}\left(\tau'\right)$. Now the variables $\tau_{1}$ and
$\tau_{2}$ are on the Keldysh contour \cite{Schwinger-Keldysh, Rammer1986} $K$ from $-\infty$ to $\infty$
and back from $\infty$ to $-\infty$.
The contour variables such as $\tau_1$ only influence the ordering of
the operators under $T_c$, and $e^{\frac{i}{\hbar}H_{0}\tau_{1}}$ has the same meaning as $e^{\frac{i}{\hbar}H_{0}t_{1}}$ with real time $t_1$.
Expanding the exponential to
perform a perturbation expansion and using Feynman diagrammatic technique,
we can obtain Dyson equations for $G^{\alpha\beta}\left(\tau_{1},\tau_{2}\right),\;\alpha,\beta=L,C,R$
such as $G_{ij}^{CL}\left(\tau_{1},\tau_{2}\right)=\sum_{l,n}\int_{K}d\tau g_{il}^{C}\left(\tau_{1},\tau\right)V_{ln}^{CL}G_{nj}^{LL}\left(\tau,\tau_{2}\right)+\sum_{l,n}\int_{K}d\tau g_{il}^{C}\left(\tau_{1},\tau\right)V_{ln}^{CR}G_{nj}^{RL}\left(\tau,\tau_{2}\right)$,
etc. All these Dyson equations could be symbolically lumped into a
compact matrix expression,
\begin{equation}
G=g+gVG=g+GVg\label{eq:dyson},
\end{equation}
where $G=\begin{bmatrix}G^{LL} & G^{LC} & G^{LR}\\
G^{CL} & G^{CC} & G^{CR}\\
G^{RL} & G^{RC} & G^{RR}
\end{bmatrix},$ $g=\begin{bmatrix}g^{L} & 0 & 0\\
0 & g^{C} & 0\\
0 & 0 & g^{R}
\end{bmatrix},$ $V=\begin{bmatrix}0 & V^{LC} & V^{LR}\\
V^{CL} & 0 & V^{CR}\\
V^{RL} & V^{RC} & 0
\end{bmatrix}$  ($V^T=V$), and
\begin{equation}
g_{jk}^{\alpha}\left(\tau_{1},\tau_{2}\right)=
-\frac{i}{\hbar}\mathrm{Tr}\left\{ \frac{e^{-\beta_{\alpha}H_{\alpha}}}{\mathrm{Tr}\left(e^{-\beta_{\alpha}H_{\alpha}}\right)}T_{c}
\left[u_{I,j}^{\alpha}\left(\tau_{1}\right)
u_{I,k}^{\alpha}\left(\tau_{2}\right)\right]\right\} ,
\end{equation}
$\alpha=L,C,R$
are equilibrium contour-ordered Green's functions for the free subsystems,
which are easy to calculate directly. No approximation
is needed here, since the coupling $H_{int}$ is quadratic.

\subsection{Generalized steady-state current formula\label{sub:Generalized-steady-state}}

Certainly, heat current flowing out of the left lead in steady state
doesn't depend on time and based on its definition $I_{L}^{ss}\equiv-\mathrm{Tr}\left[\hat{\rho}^{ss}\left(s\right){dH_{L}\left(t\right)}/{dt}\right]$ for $t>s$,
we could simply obtain
\begin{align}
I_{L}^{ss}= & -\int_{-\infty}^{\infty}\frac{d\omega}{2\pi}\,\hbar\omega\,\mathrm{Tr}\left[\left(VG^{<}\left[\omega\right]\right)_{LL}\right]\nonumber \\
= & -\int_{-\infty}^{\infty}\frac{d\omega}{2\pi}\,\hbar\omega\,\mathrm{Tr}\left[\left(V_{red}G_{red}^{<}\left[\omega\right]\right)_{LL}\right],\label{eq:ini-current}
\end{align}
where $\left(VG^{<}\left[\omega\right]\right)_{LL}$ denotes the $LL$
part submatrix of $VG^{<}\left[\omega\right]$.  Observing the
structure of $\mathrm{Tr}\left[\left(VG^{<}\left[\omega\right]\right)_{LL}\right]$,
we note that the size of the $G^{<}\left[\omega\right]$ making nonzero
contribution to $I_{L}^{ss}$ is completely determined by nonzero
entries in the symmetric total coupling matrix $V$. So we don't need
the full $G^{<}\left[\omega\right]$ which is an infinite matrix due
to the two semi-infinite leads. According to this observation, we
choose the reduced square matrix $G_{red}^{<}\left[\omega\right]$
to be the corresponding submatrix of $G^{<}\left[\omega\right]$ determined
by the row indexes of nonzero row vectors of coupling matrixes $V^{LC},V^{LR},V^{RC},V^{RL}$
plus full center part row indexes inside the total coupling matrix
$V$ for the rows of $G_{red}^{<}\left[\omega\right]$, the column
indexes of nonzero column vectors of coupling matrixes $V^{CL},V^{RL},V^{CR},V^{LR}$
plus full center part column indexes inside the total coupling matrix
$V$ for the columns of $G_{red}^{<}\left[\omega\right]$. In order
to calculate the lesser Green's function $G_{red}^{<}\left[\omega\right]$,
closed Dyson equation for reduced contour-ordered Green's function
$G_{red}\left(\tau_{1},\tau_{2}\right)$ is needed. Equation \prettyref{eq:dyson}
is the starting point and indeed it is also true that $G_{red}=g_{red}+g_{red}V_{red}G_{red},$
where $g_{red}$ is similarly defined as $G_{red}$ and $V_{red}$
is the submatrix of original $V$ after crossing out all the zero
column and row vectors except for the possible zero vectors whose
row or column indexes are the center (junction) ones. Actually, $G_{red}$
is just the corresponding submatrix of $G$ just like $V_{red}$.

From now on, for notational simplicity, we omit the
subscript $red$ of all the steady-state Green's functions and all the
coupling matrices with the understanding that these matrices are of finite dimensions.

Using the Langreth theorem \cite{Haug2008} and Fourier transforming
the obtained all kinds of real-time Green's functions, we can get
\begin{align}
 & G^{<}\left[\omega\right]\nonumber \\
= & G^{r}\left[\omega\right]\begin{pmatrix}-if_{L}\tilde{\Gamma}_{L}\left[\omega\right] & 0 & 0\\
0 & 0 & 0\\
0 & 0 & -if_{R}\tilde{\Gamma}_{R}\left[\omega\right]
\end{pmatrix}G^{a}\left[\omega\right],\label{eq:G_lesser}
\end{align}
where
\begin{equation}
\tilde{\Gamma}_{\left\{ L,R\right\} }\equiv i\left[\left(g_{\left\{ L,R\right\} }^{sur,a}\right)^{-1}-\left(g_{\left\{ L,R\right\} }^{sur,r}\right)^{-1}\right],
\end{equation}
$g_{L}^{sur,a}$ is the advanced surface Green's function for the
left lead coming from the corresponding part of the advanced reduced
Green's function $g_{red}^{a}$ and similarly for the retarded one.
This new function plays important role for our generalized Caroli formula
and for an interface formula to be derived below.
Here, fluctuation
dissipation theorem $g_{\alpha}^{<}\left[\omega\right]=f_{\alpha}\left[\omega\right]\left(g_{\alpha}^{r}-g_{\alpha}^{a}\right),\;\alpha=L,C,R.$
has been used. So is $\left(g_{C}^{a}\right)^{-1}-\left(g_{C}^{r}\right)^{-1}=0$,
which is responsible for the vanishing of junction temperature dependence
of final steady-state current formula. With respect to various
Green's functions and specific convention of Fourier transform, we
use the same definitions as Ref.~\cite{Wang2007}.

Substituting the Eq. \prettyref{eq:G_lesser} into steady current
expression \prettyref{eq:ini-current}, we can easily obtain

\begin{align}
I_{L}^{ss}= & \int_{-\infty}^{\infty}\frac{d\omega}{2\pi}\,\hbar\omega\bigl(f_{L}T_{1}\left[\omega\right]+f_{R}T_{2}\left[\omega\right]\bigr),
\end{align}
Where,
\begin{align}
 T_{1}\left[\omega\right] =i\mathrm{Tr}\left(V^{LC}G_{CL}^{r}\tilde{\Gamma}_{L}G_{LL}^{a}
 +V^{LR}G_{RL}^{r}\tilde{\Gamma}_{L}G_{LL}^{a}\right),\\
  T_{2}\left[\omega\right]
= i\mathrm{Tr}\left(V^{LC}G_{CR}^{r}\tilde{\Gamma}_{R}G_{RL}^{a}+V^{LR}G_{RR}^{r}\tilde{\Gamma}_{R}G_{RL}^{a}\right),
\end{align}
Again applying Langreth theorem and Fourier transform to the corresponding
reduced one of Eq. \prettyref{eq:dyson}, we could get $G_{LR}^{a}=G_{LL}^{a}V^{LR}g_{R}^{sur,a}+G_{LC}^{a}V^{CR}g_{R}^{sur,a}$
and $G_{LC}^{a}=G_{LL}^{a}V^{LC}g_{C}^{a}+G_{LR}^{a}V^{RC}g_{C}^{a}$.
Using the relations such as $\left(G_{CL}^{r}\right)^{\dagger}=G_{LC}^{a},\;\left(\tilde{\Gamma}_{L}\right)^{\dagger}=\tilde{\Gamma}_{L}$(where
the superscript $\dagger$ stands for transpose conjugate) etc.,
we obtain
\begin{align}
T_{1}\left[\omega\right]+T_{1}^{*}\left[\omega\right]= & \mathrm{Tr}\left(G_{RL}^{r}\tilde{\Gamma}_{L}G_{LR}^{a}\tilde{\Gamma}_{R}\right),
\end{align}
In deriving it, cyclic property of the trace was used.
Following similar steps, we could get
\begin{align}
T_{2}\left[\omega\right]+T_{2}^{*}\left[\omega\right]= & \mathrm{-Tr}\left(G_{RL}^{a}\tilde{\Gamma}_{L}G_{LR}^{r}\tilde{\Gamma}_{R}\right),
\end{align}
Due to these properties that $\tilde{\Gamma}_{\alpha}^{T}=\tilde{\Gamma}_{\alpha},\;\alpha=L,R$,
$\left(G_{RL}^{r}\right)^{T}=G_{LR}^{r}$ and $\left(G_{RL}^{a}\right)^{T}=G_{LR}^{a}$,
it is easy to show that $T_{1}+T_{1}^{*}=-\left(T_{2}+T_{2}^{*}\right).$
Now we define the general transmission coefficient
\begin{align}
T_{G}\left[\omega\right]\equiv & T_{1}\left[\omega\right]+T_{1}^{*}\left[\omega\right]=
\mathrm{Tr}\left(G_{RL}^{a}\tilde{\Gamma}_{L}G_{LR}^{r}
\tilde{\Gamma}_{R}\right).\label{eq:T_G}
\end{align}
Since current is certainly a real number, and this property has been kept in the whole derivation, we have  $I_{L}^{ss}=\frac{1}{2}\left(I_{L}^{ss}+I_{L}^{ss*}\right)=\frac{1}{2}\int_{-\infty}^{\infty}\frac{d\omega}{2\pi}\,\hbar\omega \,T_{G}\left[\omega\right]\left(f_{L}-f_{R}\right)$.

According to the definitions of retarded and advanced Green's functions
in the frequency domain, we know $G_{RL}^{a}\left[-\omega\right]=\left(G_{LR}^{r}\left[\omega\right]\right)^{T}$and
$\tilde{\Gamma}_{\left\{ L,R\right\} }\left[-\omega\right]=\left(-\tilde{\Gamma}_{\left\{ L,R\right\} }\left[\omega\right]\right)^{T}$.
Together with $f_{L}\left[-\omega\right]-f_{R}\left[-\omega\right]=-f_{L}\left[\omega\right]+f_{R}\left[\omega\right]$,
steady current $I_{L}^{ss}$ can be simplified further to the final
expression
\begin{align}
I_{L}^{ss}= & \int_{0}^{\infty}\frac{d\omega}{2\pi}\,\hbar\omega\,T_{G}\left[\omega\right]\left(f_{L}-f_{R}\right).
\end{align}
Thus, it is the same as expected that Landauer-like formula still
apply to this general case taking lead-lead interaction into account.
And this Landauer-like formula with the explicit general transmission
coefficient expression \prettyref{eq:T_G} is our central result.

Now we need to know how to calculate $G_{LR}^{r}$ in order for specific
applications. According to the corresponding reduced one of Eq. \prettyref{eq:dyson}, we can obtain a closed equation for $G_{LR}^{r}$
\begin{align}
G_{LR}^{r}= & \tilde{g}_{L}^{r}\tilde{V}^{LR,r}\tilde{g}_{R}^{r}+\tilde{g}_{L}^{r}\tilde{V}^{LR,r}\tilde{g}_{R}^{r}\tilde{V}^{RL,r}G_{LR}^{r},
\end{align}
where $\tilde{g}_{\alpha}^{r}\equiv\left(\left(g_{\alpha}^{sur,r}\right)^{-1}-V^{\alpha C}g_{C}^{r}V^{C\alpha}\right)^{-1},\;\alpha=L,R$,
and $\left(\tilde{V}^{RL,r}\right)^{T}=\tilde{V}^{LR,r}=V^{LR}+V^{LC}g_{C}^{r}V^{CR}$.
Since $G_{RL}^{a}=\left(G_{LR}^{r}\right)^{\dagger}$, now all the
quantities necessary to obtain general transmission coefficient $T_{G}$
could be expressed in terms of retarded or advanced form of submatrix
of $g_{red}$ and submatrix of $V_{red}$, which are both easily obtained.

\subsection{Recovering Caroli formula and deriving an interface formula\label{sub:Recovering-Caroli-formula}}

First, we recover Caroli formula for transmission coefficient. In
this case, coupling between the two leads $V_{LR}$ has been assumed
to be $0$. Thus, similar to what we did in subsection \prettyref{sub:Generalized-steady-state},
we could easily derived $G_{LR}^{r}=g_{L}^{sur,r}V^{LC}G_{CR}^{r}=g_{L}^{sur,r}V^{LC}G_{CC}^{r}V^{CR}g_{R}^{sur,r}$.
Together with $G_{RL}^{a}=\left(G_{LR}^{r}\right)^{\dagger}$, we
could immediately obtain from formula \prettyref{eq:T_G} that $T\left[\omega\right]=\mathrm{Tr}\left(G_{CC}^{r}\Gamma_{R}G_{CC}^{a}\Gamma_{L}\right)$.
Here, we should remember that all the quantities inside the trace
now are reduced ones. However, it is still equal to expression \prettyref{eq:Caroli formula},
in which all the quantities could be the full ones, taking trace operation and the reducing
procedure for $G$ and $V$ into account. How to calculate $T\left[\omega\right]$
and apply this efficient formula to specific applications has been
stated by many authors, e.g. \cite{Wang2008}.

Now we try to derive an interface formula still based on formula \prettyref{eq:T_G}. By interface we simply mean left lead and right lead has been connected
directly and center junction has been removed. Mathematically, we
know $V^{CL}=0$ and $V^{CR}=0$ in this situation. Consequently,
$G_{LR}^{r}=g_{L}^{sur,r}V^{LR}G_{RR}^{r}$ and $G_{RL}^{a}=\left(G_{LR}^{r}\right)^{\dagger}=G_{RR}^{a}V^{RL}g_{L}^{sur,a}$.
Straightforwardly, we get the transmission coefficient formula in
this interface problem \cite{Bijay2012}
\begin{align}
T_{I}\left[\omega\right]= & \mathrm{Tr}\left(G_{RR}^{r}\tilde{\Gamma}_{R}G_{RR}^{a}\Gamma_{L}\right).\label{eq:interface formula}
\end{align}
In order to apply this formula, still we need a closed equation for
$G_{RR}^{r}$, which could be simply obtained to be
\begin{align}
G_{RR}^{r}= & g_{R}^{sur,r}+g_{R}^{sur,r}\tilde\Sigma_{L}^{r}G_{RR}^{r},
\end{align}
where the reduced retarded self-energy is given by $\tilde\Sigma_{L}^{r}=V^{RL}g_{L}^{sur,r}V^{LR}$.

\section{AN ILLUSTRATIVE APPLICATION\label{sec:AN-ILLUSTRATIVE-APPLICATION}}

The illustrative example is a one-dimensional central ring problem,
in which there is only one particle in the center junction connected
with two semi-infinite spring chain leads. In this model, the interaction
between the two nearest particles inside the two leads also exists
taken into account as $V^{LR}$. Thus, the form of the total Hamiltonian
is the same as \prettyref{eq:total H} with $K_{0}^{\alpha},\;\alpha=L,R$
the semi-infinite tridiagonal spring constant matrix consisting of
$2\omega_{1}^{2}+\omega_{0}^{2}$ along the diagonal and $-\omega_{1}^{2}$
along the two off-diagonals, $K^{C}=2\omega_{1}^{2}+\omega_{0}^{2}$,
$V_{red}^{LC}=-\omega_{1}^{2}$, $V_{red}^{CR}=-\omega_{1}^{2}$ and
$V_{red}^{LR}=-\beta\omega_{1}^{2}$, where $\beta$ is the coupling
strength between two leads. The on-site potential term $\omega_{0}^{2}$
is necessary in establishing the steady-state current dynamically \cite{Cuansing2011}.
In this simple case, there is an analytical expression for $g_{\alpha,0}^{sur,r}\left[\omega\right],\;\alpha=L,R$,
which is $g_{\alpha,0}^{sur,r}=-{\lambda_{1}}/{\omega_{1}^{2}}$, $\lambda_{1}=(-\Omega \pm \sqrt{\Omega^{2}-4\omega_{1}^{4}})/(2\omega_{1}^{2})$,
where $\Omega=\left(\omega+i0^{+}\right)^{2}-2\omega_{1}^{2}-\omega_{0}^{2}$
and the choice between the plus or minus sign depends on satisfying
$\left|\lambda_{1}\right|<1$. And $g_{C}^{sur,r}\left[\omega\right]={1}/{\Omega}$.
After all these preparations, the transmission coefficient is simply
calculated by the formula \prettyref{eq:T_G}.

\begin{figure}
\includegraphics[width=\columnwidth]{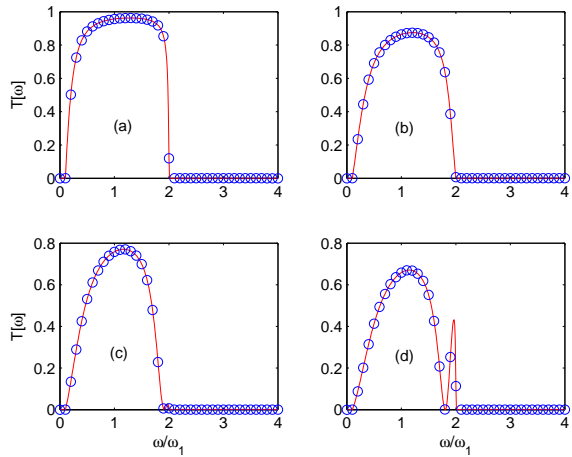}

\caption{(Color online) The transmission coefficient $T\left[\omega\right]$
as a function of frequency for coupling between leads (a) $\beta=0.2$,
(b) $\beta=0.4$, (c) $\beta=0.6$, (d) $\beta=0.8$. The results
was calculated directly (red solid line), and by repartitioning the
total Hamiltonian (blue circles). $\omega_0=0.1\,\omega_1$ in all cases. \label{fig:central ring}}

\end{figure}

Also, there is an alternative method to deal with this problem suggested
by Di Ventra as we mentioned in \prettyref{sec:INTRODUCTION}.
Essentially we repartition the total Hamiltonian so that
interaction between leads is absent. Thus, in this model,
the form of the total Hamiltonian is still the same as \prettyref{eq:total H}
but with
\begin{align}
K^{C}= & \begin{bmatrix}2\omega_{1}^{2}+\omega_{0}^{2} & -\omega_{1}^{2} & -\beta\omega_{1}^{2}\\
-\omega_{1}^{2} & 2\omega_{1}^{2}+\omega_{0}^{2} & -\omega_{1}^{2}\\
-\beta\omega_{1}^{2} & -\omega_{1}^{2} & 2\omega_{1}^{2}+\omega_{0}^{2}
\end{bmatrix},\\
V_{red}^{LC}= & \begin{bmatrix}-\omega_{1}^{2} & 0 & 0\end{bmatrix},\\
V_{red}^{CR}= & \begin{bmatrix}0\\
0\\
-\omega_{1}^{2}
\end{bmatrix}.
\end{align}
Since now $V_{red}^{LR}=0$, we can use either the Caroli formula \prettyref{eq:Caroli formula}
or the general one \prettyref{eq:T_G} to calculate the transmission
coefficient. The results of the two methods were compared in Fig.~\ref{fig:central ring}.
It turns out to be that they are the
same, which justifies the suggestion of Di Ventra from NEGF point
of view in this example.

Probably a much efficient way to calculate the transmission coefficient
in this type of noninteracting problem is to use the interface formula
\prettyref{eq:interface formula}. Frequently, the surface Green's
functions will become complex when we separate the total system into
two parts in order to apply the interface formula. However, there
are some efficient algorithms for surface Green's functions see, for
example, Ref. \cite{Wang2008}. Now we will show a specific application
of interface formula \prettyref{eq:interface formula}.

\section{an explicit interface transmission function formula}\label{sec:AN-Interface-APPLICATION}

Here in this section we derive an explicit expression for the transmission function $T_{I}[\omega]$ using Eq.~\prettyref{eq:interface formula} for the single interface setup, i.e. the left and right lead are directly connected and the center part is removed.

Let us consider that the normalized force constant for left and the right leads are $\omega_{1}^{2}$ and $\omega_{2}^{2}$ respectively and the normalized interface coupling strength is $\omega_{12}^{2}$. Also, onsite potential $\omega_{0}^{2}$ to all the atoms exists to ensure that the steady state could be established dynamically. This is a quite general scenario for a one-dimensional harmonic chain which is useful for the study of interface effects. So one of the force constant matrix say $K^{L}$ is equal to $K_{0}^{L}+\Delta K$ where $\Delta K$ is the semi-infinite matrix with only first element is nonzero $\Delta K_{11}=\omega_{12}^{2}-\omega_{1}^{2}$ while $K_{0}^{L}$ is the same as defined in the last application. Similarly for $K^{R}$ with $\omega_{1}^{2}$ replaced by $\omega_{2}^{2}$. In order to obtain the explicit form, only inputs that are required are retarded surface Green's function $g_{\alpha}^{sur,r}$ for both the leads. $G_{RR}^{\left\{r,a\right\}}$ in Eq.~\prettyref{eq:interface formula} can then be easily obtained from these expressions.

Let us calculate the surface Green's function for one of the leads, say the left lead. Then for the right lead it can be obtained just by replacing $\omega_{1}^{2}$ with $\omega_{2}^{2}$.
The surface Green's function for a semi-infinite lead when all force constants are the same is given as before, i.e. $g_{L,0}^{sur,r}$.
Now for this interface case we can obtain the surface Green's function as follows. The retarded Green's function for the left lead satisfies the following equation
\begin{equation}
\big[(\omega+i0^{+})^{2}-K^{L}\big]g_{L}^{r}=I.
\end{equation}
Taking $K^{L}=K_{0}^{L}+\Delta K$ into account, and using $\Delta K$ as a perturbation we can write
\begin{equation}
g_{L}^{r}=g_{L}^{r,0} + g_{L}^{r,0} \Delta K g_{L}^{r}.
\end{equation}
Since in this case only first atom of the left lead is connected with the first atom of the right lead, the retarded surface Green's function of the left lead is just the $(1,1)^{\rm th}$ element of $g_{L}^{r}$ and we obtain
\begin{equation}
g_{L}^{sur,r}=\frac{1}{\omega_{1}^{2}-\omega_{12}^{2}-\omega_{1}^{2}/\lambda_{1}},
\end{equation}
and the self-energy for the lead is given by
\begin{equation}
\tilde\Sigma_{L}^{r}=\frac{\omega_{12}^{4}}{\omega_{1}^{2}-\omega_{12}^{2}-\omega_{1}^{2}/\lambda_{1}}.
\end{equation}
Knowing this surface Green's function and self-energy, we can easily obtain ${T_{I}}[\omega]$ from Eq.~\prettyref{eq:interface formula} which can be written as
\begin{equation}
{T_{I}}[\omega]= -\frac{\omega_{1}^{2} \omega_{2}^{2} \omega_{12}^{4} (\lambda_1-\lambda_1^{*})(\lambda_2-\lambda_2^{*})}{\big|(\omega_{1}^{2}-\omega_{12}^{2}-\omega_{1}^{2}/\lambda_1)(\omega_{2}^{2}-\omega_{12}^{2}-\omega_{2}^{2}/\lambda_2)-\omega_{12}^{4}\big|^{2}},
\end{equation}
where, $\lambda_2$ is similarly defined as $\lambda_1$ with $\omega_1$ replaced by $\omega_2$.
It has been noted that it matches exactly with the result in Ref. \cite{Zhang2011}, where this expression is obtained from wave-scattering method.
Now if $\omega_{1}^{2}=\omega_{2}^{2}=\omega_{12}^{2}$ then we have perfect transmission i.e., ${T_{I}}[\omega]=1$ for $\omega$ within the phonon band $\omega_{0}^{2} \le \omega^{2} \le 4\omega_{1}^{2}+\omega_{0}^{2}$ and $0$ outside this region.

\section{SUMMARY}

We examine the heat current in a lead-junction-lead quantum system,
in which coupling between the leads has been taken into account. After
assuming ideal steady state could be established from initial product
state, we rigorously derived a general Landauer-like formula in the NEGF framework,
from which the corresponding transmission coefficient was obtained.
Based on this general transmission coefficient formula, Caroli formula
was recovered and a computationally efficient interface formula applicable
to the case in which the total noninteracting Hamiltonian could be
repartitioned was derived. Also an illustrative example was given
as both a verification of the validity of the repartitioning procedure
which doesn't affect the steady current value and the clarification
of the meaning of some quantities used in the formula such as $V_{red}^{LC}$
etc in different situations. Finally, we derived an explicit transmission coefficient formula in
a quite general one-dimensional interface situation based on interface formula,
which turned out to be perfectly consistent with result obtained by wave-scattering method.

\begin{acknowledgments}
We would like to thank Lifa Zhang and Juzar Thingna for insightful
discussions.  This work is supported in part by a URC grant R-144-000-257-112.

\end{acknowledgments}


\begin{thebibliography}{99}
\bibitem{Rego1998} L. G. C. Rego and G. Kirczenow, Phys. Rev. Lett. \textbf{81}, 232, (1998).
\bibitem{Blencowe1999} M. P. Blencowe, Phys. Rev. B \textbf{59}, 4992 (1999).
\bibitem{Segal2003} D. Segal, A. Nitzan, and P. H\"anggi, J. Chem. Phys. \textbf{119}, 6840 (2003).
\bibitem{Dhar2006} A. Dhar and D. Roy, J. Stat. Phys. \textbf{125}, 805 (2006).
\bibitem{Dhar2008}A. Dhar, Adv. in Phys., \textbf{57}, 457-537 (2008).
\bibitem{Ozpineci2001} A.Ozpineci and S. Ciraci, Phys. Rev. B \textbf{63}, 125415 (2001).
\bibitem{Wang2006} J.-S. Wang, J. Wang, and N. Zeng, Phys. Rev. B \textbf{74}, 033408 (2006).
\bibitem{Galperin2007} M. Galperin, A. Nitzan, and M. A. Ratner,   Phys. Rev. B  \textbf{75}, 155312 (2007).
\bibitem{Meir1992} Y. Meir and N. S. Wingreen, Phys. Rev. Lett. \textbf{68}, 2512 (1992).
\bibitem{Yamamoto2006} T. Yamamoto and K. Watanabe, Phys. Rev. Lett. \textbf{96}, 255503 (2006).
\bibitem{Caroli1971} C. Caroli, R. Combescot, P. Nozieres, D. Saint-James, J. Phys. C: Solid St. Phys. \textbf{4}, 916 (1971).
\bibitem{Mingo2003} N. Mingo, L. Yang, Phys. Rev. B \textbf{68}, 245406, (2003).
\bibitem{Zhang2007} W. Zhang, T. S. Fisher, and N. Mingo, Numer. Heat Transf. Part B, \textbf{51}, 333 (2007).
\bibitem{Suman2012} S. G. Das and A. Dhar  arXiv:1204.5595.
\bibitem{Ventra2008} M. Di. Ventra, \textit{Electrical Transport in Nanoscale Systems}, Cambridge University Press, 2008.
\bibitem{Rubin1971} R. J. Rubin and W. L. Greer, J. Math. Phys. {\bf 12}, 1686 (1971).
\bibitem{Bijay2011}J.-S. Wang, B. K. Agarwalla, and H. Li, Phys. Rev. B \textbf{84}, 153412, (2011).
\bibitem{Ness2010} H. Ness, L. K. Dash, and R. W. Godby, Phys. Rev. B \textbf{82}, 085426 (2010).
\bibitem{Xu2008}Y. Xu, J.-S. Wang, W. Duan, B.-L. Gu, and B. Li, Phys. Rev. B {78}, 224303 (2008).
\bibitem{Haug2008} H. Haug and A.-P. Jauho, \textit{Quantum Kinetics in Transport and Optics of Semiconductors}, 2nd ed. (Springer, New York, 2008).
\bibitem{Cuansing2011} E. C. Cuansing, H. Li, and J.-S. Wang, arXiv:1105.2233.
\bibitem{Schwinger-Keldysh} J. Schwinger, J. Math. Phys. {\bf 2}, 407
  (1961); L.V. Keldysh, Sov. Phys. JETP {\bf 20}, 1018 (1965).
\bibitem{Rammer1986} J. Rammer and H. Smith, Rev. Mod.
  Phys. {\bf 58}, 323 (1986).
\bibitem{Wang2007} J.-S. Wang, N. Zeng, J. Wang, and C. K. Gan, Phys. Rev. E \textbf{75}, 061128 (2007).
\bibitem{Wang2008} J.-S. Wang, J. Wang, and J. T. L\"u,
Eur. Phys. J. B \textbf{62}, 381 (2008).
\bibitem{Bijay2012} B. K. Agarwalla, J.-S. Wang, and B. Li, Unpublished.
\bibitem{Zhang2011} L. Zhang, P. Keblinski, J.-S. Wang, and B. Li, Phys. Rev. B \textbf{83}, 064303 (2011).













\end{thebibliography}

\end{document}